\documentclass[10pt,conference]{IEEEtran}
\IEEEoverridecommandlockouts

\usepackage[T1]{fontenc}
\usepackage{cite}
\usepackage{amsmath,amssymb,amsfonts}
\usepackage{graphicx}
\usepackage{textcomp}
\usepackage{xcolor}
\usepackage{booktabs}
\usepackage[group-separator={,},group-minimum-digits=4]{siunitx}
\usepackage{pgfplots}
\pgfplotsset{compat=1.17}
\usepackage{listings}
\usepackage[hidelinks]{hyperref}
\usepackage{url}

\newcommand{\BenchN}{236}
\newcommand{\SourceN}{237}
\newcommand{\BenchROSn}{103}
\newcommand{\BenchRepos}{28}
\newcommand{\BenchROSRepos}{17}
\newcommand{\BenchDn}{38}
\newcommand{\RemovedN}{9}
\newcommand{\ResRecentN}{153}
\newcommand{\RefLinesMedian}{30}
\newcommand{\RefFilesMedian}{2}
\newcommand{\PrimaryAttempts}{3,540}
\newcommand{\Runso}{3}
\newcommand{\ResoPct}{56.5}
\newcommand{\ResoLo}{50.7}
\newcommand{\ResoHi}{62.4}
\newcommand{\Anyo}{60.2}
\newcommand{\Everyo}{51.7}
\newcommand{\CostoMean}{0.47}
\newcommand{\Runsg}{3}
\newcommand{\ResgPct}{52.5}
\newcommand{\ResgLo}{46.5}
\newcommand{\ResgHi}{58.6}
\newcommand{\Anyg}{56.4}
\newcommand{\Everyg}{47.9}
\newcommand{\CostgMean}{0.23}
\newcommand{\Runsh}{3}
\newcommand{\ReshPct}{54.1}
\newcommand{\ReshLo}{47.9}
\newcommand{\ReshHi}{60.3}
\newcommand{\Anyh}{57.2}
\newcommand{\Everyh}{50.0}
\newcommand{\CosthMean}{0.43}
\newcommand{\Runsa}{3}
\newcommand{\ResaPct}{51.4}
\newcommand{\ResaLo}{45.1}
\newcommand{\ResaHi}{57.9}
\newcommand{\Anya}{54.2}
\newcommand{\Everya}{47.9}
\newcommand{\CostaMean}{1.24}
\newcommand{\Runss}{3}
\newcommand{\RessPct}{45.8}
\newcommand{\RessLo}{39.8}
\newcommand{\RessHi}{51.7}
\newcommand{\Anys}{51.3}
\newcommand{\Everys}{39.4}
\newcommand{\CostsMean}{0.10}
\newcommand{\Runsx}{1}
\newcommand{\ResxPct}{46.2}
\newcommand{\ResxLo}{39.9}
\newcommand{\ResxHi}{52.6}
\newcommand{\Anyx}{--}
\newcommand{\Everyx}{--}
\newcommand{\CostxMean}{--}
\newcommand{\Runsk}{2}
\newcommand{\ReskPct}{43.9}
\newcommand{\ReskLo}{37.5}
\newcommand{\ReskHi}{50.0}
\newcommand{\Anyk}{47.0}
\newcommand{\Everyk}{40.7}
\newcommand{\CostkMean}{--}
\newcommand{\ResROSo}{48.5}
\newcommand{\ResROSg}{49.5}
\newcommand{\ResROSh}{50.8}
\newcommand{\ResROSa}{46.3}
\newcommand{\ResROSs}{39.8}
\newcommand{\ResROSx}{42.7}
\newcommand{\ResROSk}{39.3}
\newcommand{\ResBo}{59.8}
\newcommand{\ResBg}{55.0}
\newcommand{\ResBh}{55.6}
\newcommand{\ResBa}{56.1}
\newcommand{\ResBs}{55.6}
\newcommand{\ResBx}{52.4}
\newcommand{\ResBk}{54.0}
\newcommand{\ResCo}{69.8}
\newcommand{\ResCg}{61.5}
\newcommand{\ResCh}{64.6}
\newcommand{\ResCa}{59.4}
\newcommand{\ResCs}{49.0}
\newcommand{\ResCx}{53.1}
\newcommand{\ResCk}{46.9}
\newcommand{\ResDo}{61.4}
\newcommand{\ResDg}{49.1}
\newcommand{\ResDh}{51.8}
\newcommand{\ResDa}{50.9}
\newcommand{\ResDs}{43.0}
\newcommand{\ResDx}{39.5}
\newcommand{\ResDk}{36.8}
\newcommand{\BinSmallN}{78}
\newcommand{\BinLargeN}{24}
\newcommand{\CatRunsN}{15}
\newcommand{\CatAlways}{83}
\newcommand{\CatNever}{81}
\newcommand{\CensusCandidates}{61,878}
\newcommand{\CensusROSTwo}{2,114}
\newcommand{\CensusNew}{2,043}
\newcommand{\CensusActive}{197}
\newcommand{\CensusRobot}{171}
\newcommand{\CensusFrame}{80}
\newcommand{\CuratedN}{42}
\newcommand{\CuratedKept}{25}
\newcommand{\ROSReposCand}{39}
\newcommand{\RejectNoCompile}{96}
\newcommand{\NotReleasedN}{6}
\newcommand{\FunnelTrackACandidates}{1,567}
\newcommand{\FunnelBuildAttempted}{928}
\newcommand{\FunnelEnvironmentBuilt}{489}
\newcommand{\FunnelAdmitted}{262}
\newcommand{\FunnelPRs}{61,655}
\newcommand{\StmtRemovedN}{10}
\newcommand{\ExId}{ros2\_\_rosbag2-1812}
\newcommand{\ExRepo}{ros2/rosbag2}
\newcommand{\ExPR}{1812}
\newcommand{\ExIssueWords}{234}
\newcommand{\ExFiles}{2}
\newcommand{\ExLines}{11}
\newcommand{\ExFtpN}{1}
\newcommand{\ExPtpN}{7}
\newcommand{\ExFtp}{test\_\allowbreak{}local\_\allowbreak{}message\_\allowbreak{}definition\_\allowbreak{}source.\allowbreak{}throw\_\allowbreak{}definition\_\allowbreak{}not\_\allowbreak{}found\_\allowbreak{}for\_\allowbreak{}unknown\_\allowbreak{}msg}
\newcommand{\ExMerged}{2024-09-23}
\newcommand{\HiddenN}{26}
\newcommand{\MultiIssueN}{11}
\newcommand{\TrajAvail}{3,529}
\newcommand{\CapacityReplacements}{14}
\newcommand{\CppRosN}{80}
\newcommand{\DataDateUS}{November 10, 2026}
\newcommand{\MixedN}{72}
\newcommand{\MjlabN}{34}
\newcommand{\PostponedN}{229}
\newcommand{\PostponedSimN}{192}
\newcommand{\RecentPct}{65}
\newcommand{\RepoAlg}{5}
\newcommand{\RepoCandAlg}{11}
\newcommand{\RepoCandAll}{63}
\newcommand{\RepoCandLearn}{2}
\newcommand{\RepoCandSim}{11}
\newcommand{\RepoLearn}{2}
\newcommand{\RepoSim}{4}
\newcommand{\SingleTaskRepos}{9}
\newcommand{\TopFiveN}{133}
\newcommand{\TopFivePct}{56}
\newcommand{\UnprocessedN}{220}
\newcommand{\WithheldN}{26}
\definecolor{copus}{HTML}{1F4E79}
\definecolor{csonnet}{HTML}{7FB3D5}
\definecolor{csol}{HTML}{C55A11}
\definecolor{cmax}{HTML}{6B4C9A}
\definecolor{cglm}{HTML}{4B555D}
\definecolor{castra}{HTML}{F4B183}
\definecolor{ckimi}{HTML}{8C8C8C}
\definecolor{cros}{HTML}{1B9E77}
\definecolor{csim}{HTML}{D95F02}
\definecolor{calg}{HTML}{7570B3}
\definecolor{clearn}{HTML}{E7298A}
\definecolor{calways}{HTML}{1B9E77}
\definecolor{cflips}{HTML}{FDB863}
\definecolor{csplit}{HTML}{7570B3}
\definecolor{cnever}{HTML}{636363}
\pgfplotsset{
  every axis/.append style={font=\scriptsize, tick label style={font=\scriptsize}, label style={font=\scriptsize},
    legend style={font=\scriptsize, draw=none, fill=none}, axis line style={gray!60}, tick style={gray!60}},
  paperbar/.style={draw=none, fill=#1},
}
\lstdefinestyle{diff}{moredelim=[is][\color{green!45!black}]{|+|}{|/|}, moredelim=[is][\color{red!70!black}]{|-|}{|/|}}

\newcommand{\toolname}{\textsc{robo-swe}}
\newcommand{\toolurl}{\url{https://evalium.ai/robo-swe}}
\newcommand{\dataurl}{\url{https://huggingface.co/datasets/vharatian/Robo-SWE}}

\begin{document}

\title{\toolname: An Executable Issue-Resolution\\Dataset Mined from Robot Software Repositories%
\thanks{Vahid Haratian and Vahid Ansari contributed equally to this work.}}

\author{
\IEEEauthorblockN{
Vahid Haratian\IEEEauthorrefmark{1}\IEEEauthorrefmark{2},
Vahid Ansari\IEEEauthorrefmark{2},
Mohammadhossein Malmir\IEEEauthorrefmark{3},
Alois Knoll\IEEEauthorrefmark{3}}
\IEEEauthorblockA{\IEEEauthorrefmark{1}Bilkent University, Ankara, Turkey}
\IEEEauthorblockA{\IEEEauthorrefmark{2}Evalium, Yerevan, Armenia}
\IEEEauthorblockA{\IEEEauthorrefmark{3}Technical University of Munich, Munich, Germany}
\IEEEauthorblockA{\{vahid.haratian,vahid.ansari\}@evalium.ai,
\{hossein.malmir,k\}@tum.de}
}

\maketitle

\begin{abstract}
Coding agents are evaluated mostly on general-purpose software, but domain-specific systems require
domain-aware evaluation. Robot software is a particularly challenging case: Robot Operating System~2
(ROS~2) packages have their own build tools, simulators and learning frameworks are built on physics
engines, and behavior that depends on timing and hardware is hard to test. We present \toolname, a
dataset of \BenchN{} executable issue-resolution tasks mined from merged pull requests in \BenchRepos{}
public robotics repositories in four domains: ROS~2 systems, simulation and physics engines, robotics
algorithm libraries, and robot-learning frameworks. Each task contains the original issue text, a container recipe
for the repository at the commit before the fix, the developers' fix, and the developers' tests,
which fail before the fix and pass after it in repeated offline runs. The dataset also keeps a provenance
ledger of all \FunnelTrackACandidates{} candidate pull requests, with the stage each reached and the reason
for every rejection, and, as a first use, \PrimaryAttempts{} graded attempts by five coding-agent setups
with recorded patches and \TrajAvail{} available trajectories. Across three runs per primary setup,
resolve rates range from \RessPct\% for Claude Sonnet~5.5 to \ResoPct\% for Claude Opus~5.5; GPT-6.1 Sol
reaches \ReshPct\% at maximum effort and \ResgPct\% at medium effort, and GPT-6 Astra \ResaPct\%.
GLM-5.3 resolves \ResxPct\%, and Kimi K3.1 Pro \ReskPct\%. The tasks are validated by execution; verification of their fairness by human reviewers
is future work. The dataset and public leaderboard are available online.
\end{abstract}


\section{Introduction}

Robot software consists of large, interdependent stacks of reusable components~\cite{kolak2020village}.
A mobile manipulator built on the Robot Operating System (ROS), in its current generation
ROS~2~\cite{macenski2022ros2}, combines navigation, control, perception, hardware
drivers, and middleware across C++ and Python packages built with colcon and CMake. Development
depends on simulators and physics engines, while robot-learning frameworks add substantial codebases for
policies trained in simulation before deployment to hardware~\cite{zakka2025playground,josifovski2024cdr}. Maintaining these
stacks is central to robotics engineering, yet testing them is difficult: behavior depends on timing,
concurrency, and hardware, and practitioners still rely heavily on field testing~\cite{afzal2020challenges}.

These characteristics distinguish robot-software repair from repair in self-contained applications.
Correctness spans packages and processes: a patch may compile yet violate a ROS message or lifecycle
contract, change launch-time behavior, or fail only with a particular simulator, middleware, timing, or
hardware configuration. Reproducing a task therefore requires more than checking out source and running
unit tests. Evaluation must preserve the domain toolchain and execution environment, account for
concurrency and nondeterminism, and separate infrastructure failures from incorrect repairs; otherwise,
scores may measure environment mismatch rather than an agent's repair ability.

Despite these domain constraints, coding agents are typically evaluated by whether they resolve
GitHub issues in general-purpose repositories. SWE-bench~\cite{jimenez2024swebench}
and its human-checked subset~\cite{chowdhury2024verified} defined the task; later datasets extended it to
more languages and automated its
collection~\cite{zan2025multiswe,zhang2025swebenchlive,badertdinov2025swerebench}. None of them targets robot
software or its build systems. Roboticists, meanwhile, ask whether such agents can supply the
engineered, testable modules that robot learning does not obtain from
data~\cite{goldberg2025datagap,goldberg2026plenary}. Current robotics benchmarks do not answer this for
existing codebases: agents write new programs against a robot application programming interface (API) or a
simulator~\cite{liang2023cap,fu2026capx,shen2026embodiedswe,ma2026rlebench}, complete single
functions~\cite{zhao2027rosdeveval}, or repair a small curated defect set in one
stack~\cite{tuzikov2026robobugbench}.

We present \toolname, a dataset mined from the history of robotics projects to close this gap. It is meant
for software engineering researchers who study or train coding agents, and for robotics teams who want to
know which parts of their stack an agent can maintain. It provides:
\begin{itemize}
\item \BenchN{} executable issue-resolution tasks from \BenchRepos{} repositories in four domains
(Table~\ref{tab:funnel}), among them Navigation2, ros2\_control, rosbag2, MoveIt~2, Newton, MuJoCo Warp,
python-control, and LeRobot. Each pairs the original issue with a pre-fix workspace, the developers' fix, and developer tests that fail before
the fix and pass after it.
\item The provenance of the tasks: a 2026 refresh of a census of ROS repositories on
GitHub~\cite{malavolta2020architect}, and a ledger of all \FunnelTrackACandidates{} candidate pull requests
with the stage that each reached and the reason for every rejection.
\item A first use: \PrimaryAttempts{} graded attempts by five coding-agent setups, with their patches,
per-test verdicts, and trajectories.
\end{itemize}
Dataset: \dataurl.
Leaderboard: \toolurl.

\section{Dataset Construction}
\label{sec:construction}

One pipeline builds all four domains with the same task definition, mining window, statement, and export
format; Table~\ref{tab:funnel} follows the candidates through it. A task is a merged pull request (PR) that
closes an earlier issue and changes production code together with tests, so that the developers' own tests
can serve as the oracle. Its author must be human, and it must not be a backport.

\begin{table}[t]
\caption{Candidate filtering by domain, from \FunnelTrackACandidates{} distinct pull requests to
\BenchN{} evaluated tasks.}
\label{tab:funnel}
\centering\footnotesize
\setlength{\tabcolsep}{4pt}
\begin{tabular}{@{}lrrrrr@{}}
\toprule
 & ROS~2 & Sim. & Alg. & Learn. & All \\
\midrule
Candidate PRs & \num{640} & \num{709} & \num{140} & \num{78} & \num{1567} \\
Queued for building & \num{325} & \num{458} & \num{93} & \num{52} & \num{928} \\
Environment built & \num{291} & \num{104} & \num{50} & \num{44} & \num{489} \\
Admitted by execution & \num{124} & \num{66} & \num{34} & \num{38} & \num{262} \\
\textbf{Evaluated tasks} & \num{103} & \num{63} & \num{32} & \num{38} & \num{236} \\
\midrule
Repositories with candidates & \ROSReposCand & \RepoCandSim & \RepoCandAlg & \RepoCandLearn & \RepoCandAll \\
Repositories with evaluated tasks & \BenchROSRepos & \RepoSim & \RepoAlg & \RepoLearn & \BenchRepos \\
Resolved in every baseline run$^{\ast}$ & 29 & 31 & 14 & 9 & \CatAlways \\
Resolved in no baseline run$^{\ast}$ & 39 & 21 & 7 & 14 & \CatNever \\
\bottomrule
\multicolumn{6}{@{}l}{\scriptsize $^{\ast}$\,\CatRunsN{} runs of the five primary setups (Section~\ref{sec:baseline}).}
\end{tabular}
\end{table}

Build-queue counts include postponements; ``environment built'' denotes a successful recorded build.
Release checks remove \WithheldN{} admitted candidates independently of agent scores. Sim., Alg., and
Learn. denote simulation and physics, robotics algorithms, and robot learning, respectively.

\subsection{Repositories}

\emph{ROS~2.} The census of ROS-based systems by Malavolta et
al.~\cite{malavolta2020architect,malavolta2021mining} uses a 2019 snapshot. A recent survival study samples
810 ROS~2 repositories from the ROS Index~\cite{freitas2026walking}, for a different research purpose. We re-implemented the census
mining on the GitHub GraphQL API with the same thresholds, extended the keyword lists with ROS~2 terms, and
added the rosdistro index and community lists (September 21--28, 2026). Of \CensusCandidates{} candidate
repositories, \CensusROSTwo{} pass the original filters and contain ROS~2 code; \CensusNew{} of them are not
in the 2019 list. Of the \CensusActive{} that are active and have at least 100 merged PRs, GPT-5 mini
classifies \CensusRobot{} as robot systems. The \CensusFrame{} largest with at least five human authors,
none holding more than 80\% of the PRs, form the mining frame.

\emph{Simulation and physics, robotics algorithms, and robot learning.} These three domains use a
curated list of \CuratedN{} repositories and keep the \CuratedKept{} that are active, have at least 30 merged PRs from five or more
human authors in the window, merge through PRs, and carry an open-source license.

\subsection{Candidate Mining and Screening}

We mine a four-year window ending in October 2026: for ROS~2, a stratified sample of up to 500
PRs from the last two years of each of the \CensusFrame{} repositories, and then every PR in the window for
the \ROSReposCand{} repositories that yielded a candidate; for simulation, algorithms, and learning, every PR of the
\CuratedKept{} repositories. Production and test changes are told apart by content, not by file name, and
size limits keep tasks reviewable (ROS~2: at most 25 changed files; elsewhere: at most 100 files, 10 of them
production files). The \FunnelPRs{} merged PRs yield \FunnelTrackACandidates{} candidates.

Large language model (LLM) screening assesses issue clarity and whether developer tests plausibly
check the issue without requiring a fix-only API~\cite{aleithan2024swebenchplus}. ROS~2 screening uses GPT-5
with issue text, PR metadata, paths, and test identifiers, without the patch. The other three
domains use GPT-5 mini (2025-08-07) with issue evidence and developer test changes; a deterministic gate checks the
recorded facts. These checks select candidates for building; execution decides admission. Final task
statements use the original issue titles and bodies; \MultiIssueN{} tasks close multiple issues.

\subsection{Environments and Execution-Based Admission}

Each candidate gets a container at the PR's base commit, and its tests run without network access. ROS~2
images start from a digest-pinned ROS image (with a dated package snapshot for older commits), install
dependencies with rosdep, and build the touched packages with colcon; the distribution follows the PR's
target branch or the continuous-integration configuration at the base commit. The other domains use one
reviewed recipe per repository with pinned base images and package versions. Environments were built for
\FunnelEnvironmentBuilt{} of the \FunnelBuildAttempted{} queued candidates. The simulation, algorithm, and learning queues were not exhausted: \PostponedN{} candidates
(\PostponedSimN{} in simulation) remained marked postponed for disk space, including one with an earlier
successful build; \UnprocessedN{} more had no build attempt. Thus, queue size minus successful builds is
not a count of technical build failures.

Admission runs the tests in three states: (A) the base commit with its own tests (outside ROS~2 only);
(B) the base commit with the PR's tests, where the \emph{fail-to-pass} (F2P) tests must fail; and (C)
state~B plus the developers' fix, where the F2P tests must pass and no \emph{pass-to-pass} (P2P) test may
break. Each state is repeated two or three times, and unstable verdicts are never admitted: in ROS~2, a
flaky test is dropped, and a task without a stable F2P test is rejected; elsewhere, any flaky verdict rejects
the task. A test that does not compile before the fix does not count as failing. This rule alone rejects
\RejectNoCompile{} ROS~2 candidates, an obstacle particular to compiled languages, where new tests often
call the API that the fix introduces. Tasks are never edited to pass. In total, \FunnelAdmitted{}
candidates are admitted.

\subsection{Release Checks}

Admitted tasks are exported in the Harbor task format~\cite{merrill2026terminalbench} and checked again by
the harness: the developers' fix must score~1, and an empty patch~0. Of the admitted tasks, \WithheldN{}
were withheld, each with a recorded reason: \NotReleasedN{} at export (two flaky or already passing at the
base, three whose issue contains the fix, and one whose tests need an API the issue never mentions); \RemovedN{}
after the harness check or inspection (two whose F2P tests pass without any change, four whose fix fails in
the exported image, one whose environment no longer builds, and two whose hidden tests overwrite a build
file that an agent may legitimately edit); \StmtRemovedN{} ROS~2 tasks rejected by a final LLM check of the
statement with GPT-5 mini (two contain the fix, six have tests beyond what the issue asks, and two do not pose a
well-defined task); and one whose pinned nightly dependency was retired upstream. The original \SourceN{}-task frozen
release is retained; this last task is excluded uniformly from every setup, giving \BenchN{} evaluated
tasks. Several ROS~2 test suites are sensitive to host load: with 14 tasks graded
concurrently on 12 cores, the developers' fix failed on 15 ROS~2 tasks, 11 of which pass on a quiet host.

\section{Dataset Description}
\label{sec:dataset}

\subsection{Contents}

Table~\ref{tab:funnel} gives the size of each domain. The ROS~2 tasks cover navigation and mobile robots
(45 tasks), control and drivers (26), middleware and infrastructure (22), task planning (5), manipulation
(4), and perception (1); the largest sources are Navigation2, rosbag2, and ros2\_control. Outside ROS~2, the tasks come from simulators and physics engines that run on graphics processing units (GPUs), such as Newton
and MuJoCo Warp, from algorithm libraries (python-control, Robotics Toolbox), and from robot-learning
frameworks (mjlab, LeRobot). The fix touches C++ in \CppRosN{} of the \BenchROSn{} ROS~2 tasks and is pure
Python in every task outside ROS~2: no C++ task from the curated simulation, algorithm, or learning projects is in this evaluation. The developers' fixes are small: a median of \RefLinesMedian{} changed lines in
\RefFilesMedian{} files; \BinSmallN{} tasks change at most 15 lines and \BinLargeN{} more than 150. The PRs
were merged between November 2022 and October 2026; \ResRecentN{} (\RecentPct\%) were merged in 2026, and
every task records its merge date, so that users can exclude tasks that predate a model's training cutoff.

\begin{figure}[t]
\centering
\fbox{\begin{minipage}{\dimexpr\columnwidth-2\fboxsep-2\fboxrule\relax}\scriptsize\raggedright
\textbf{\texttt{\ExId/}}\hfill \ExRepo{} PR~\#\ExPR, merged \ExMerged\\[2pt]
\texttt{instruction.md}\quad the issue as filed (\ExIssueWords{} words), then the workspace path and the build command
\begin{lstlisting}
`ros2 bag convert` ignoring messages with unknown types
The `ros2 bag convert` fails to convert bag files with "unknown" message types. Unknown is the type which typesupport is not available during the bag conversion.
\end{lstlisting}
\texttt{environment/}\quad Dockerfile: digest-pinned ROS~2 image, rosdep, colcon build at the base commit\\[2pt]
\texttt{solution/}\quad the developers' fix (\ExLines{} changed lines in \ExFiles{} files), hidden from the agent
\begin{lstlisting}[style=diff]
rosbag2_cpp/src/rosbag2_cpp/message_definitions/local_message_definition_source.cpp @@ -179,7 +180,13 @@
   std::string package = match[1];
|-|-  std::string share_dir = ament_index_cpp::get_package_share_directory(package);|/|
|+|+  std::string share_dir;|/|
|+|+  try {|/|
|+|+    share_dir = ament_index_cpp::get_package_share_directory(package);|/|
|+|+  } catch (const ament_index_cpp::PackageNotFoundError & e) {|/|
|+|+    ROSBAG2_CPP_LOG_WARN("'%s'", e.what());|/|
|+|+    throw DefinitionNotFoundError(definition_identifier.topic_type());|/|
|+|+  }|/|
\end{lstlisting}
\texttt{tests/}\quad hidden oracle: the PR's test changes, \ExFtpN{} F2P and \ExPtpN{} P2P test identifiers, the grading
script. F2P: \texttt{\ExFtp}\\[2pt]
\texttt{task.toml}\quad time, processor, and memory limits; network mode; provenance
\end{minipage}}
\caption{\textbf{A \toolname{} task and its files.} The agent sees \texttt{instruction.md} and the built
workspace; \texttt{solution/} and \texttt{tests/} stay hidden until grading.}
\label{fig:task}
\end{figure}

\subsection{Storage and Schema}

Each task is a directory in the Harbor task format (Fig.~\ref{fig:task}). The working pipeline and review
application store metadata in SQLite ledgers. The replication package organizes four tables,
with comma-separated values (CSV) and JSON Lines exports: \emph{tasks}, one
record per task with repository, PR and issue links, base and fix commits, merge date, domain and
subcategory, build system, test framework, patch size, F2P and P2P test identifiers, license, content hash,
and the audit flag of Section~\ref{sec:limits}; \emph{candidates}, one record per candidate PR with the last
stage it reached and the reason it was dropped; \emph{admission}, the verdict of every test in every
repetition of states A--C; and \emph{attempts}, one record per agent attempt (Section~\ref{sec:baseline})
with setup, run, outcome, per-test verdicts, token use, time, and the paths of the agent's patch and
trajectory.
Task versions and the uniform environment exclusion must be recorded with each score.

\subsection{Usage}

An agent is run on a task with \texttt{harbor run -p <task> -a <agent> -m <model>}. It works in the built
workspace at the base commit with the tests withheld; the hidden tests are then copied in and executed. A
task is \emph{resolved} when every F2P test passes and no P2P test fails. Because the fixes are public,
agents should run without web access; because of the load sensitivity above, we grade at most three ROS~2
tasks at a time on a 12-core host.

\section{Baseline Evaluation}
\label{sec:baseline}

As a first use and a reference point, we ran five agent \emph{setups} (a setup fixes the model, the agent
harness, and the reasoning effort) three times over all tasks, \PrimaryAttempts{} graded attempts: Claude
Opus~5.5 and Claude Sonnet~5.5 in Claude Code~2.1.302, and GPT-6.1 Sol (medium and max effort) and GPT-6
Astra in Codex~0.163.0. Two further setups in OpenCode, GLM-5.3 and Kimi K3.1 Pro, have \Runsx{} and
\Runsk{} complete runs. Each attempt receives the issue text and the built workspace, has a one-hour
budget, and reaches only the model's API; the Claude Code setups also have a cost cap of \$20 per attempt.
Provider interruptions are excluded rather than scored as model failures; \CapacityReplacements{}
capacity-interrupted primary attempts were replaced, including four with passing raw rewards. Ordinary
graded failures retain their scores. Only complete runs are included in Table~\ref{tab:results}.

\begin{table}[t]
\caption{Task-resolution rates by model--agent setup and domain (\DataDateUS).}
\label{tab:results}
\centering\scriptsize
\setlength{\tabcolsep}{2.3pt}
\begin{tabular}{@{}lccccccccc@{}}
\toprule
Setup (runs) & Total & Interval & ROS\,2 & Sim. & Alg. & Learn. & Any & Every & \$ \\
\midrule
Opus~5.5 (\Runso) & \ResoPct & \ResoLo--\ResoHi & \ResROSo & \ResBo & \ResCo & \ResDo & \Anyo & \Everyo & \CostoMean \\
Sol max (\Runsh) & \ReshPct & \ReshLo--\ReshHi & \ResROSh & \ResBh & \ResCh & \ResDh & \Anyh & \Everyh & \CosthMean \\
Sol medium (\Runsg) & \ResgPct & \ResgLo--\ResgHi & \ResROSg & \ResBg & \ResCg & \ResDg & \Anyg & \Everyg & \CostgMean \\
Astra (\Runsa) & \ResaPct & \ResaLo--\ResaHi & \ResROSa & \ResBa & \ResCa & \ResDa & \Anya & \Everya & \CostaMean \\
Sonnet~5.5 (\Runss) & \RessPct & \RessLo--\RessHi & \ResROSs & \ResBs & \ResCs & \ResDs & \Anys & \Everys & \CostsMean \\
\midrule
GLM-5.3 (\Runsx) & \ResxPct & \ResxLo--\ResxHi & \ResROSx & \ResBx & \ResCx & \ResDx & \Anyx & \Everyx & \CostxMean \\
Kimi K3.1 Pro (\Runsk) & \ReskPct & \ReskLo--\ReskHi & \ResROSk & \ResBk & \ResCk & \ResDk & \Anyk & \Everyk & \CostkMean \\
\bottomrule
\end{tabular}
\end{table}

Table~\ref{tab:results} averages task-level resolve rates over complete runs. Any and Every denote tasks
resolved in at least one and in every run, respectively. The cost column reports mean API-equivalent cost
per graded attempt in complete runs; -- denotes unavailable data. The 95\% intervals use 2,000
task-bootstrap resamples (Wilson when task scores are binary). Across the \CatRunsN{} primary
setup-runs, \CatAlways{} tasks are always resolved, \CatNever{} never, and \MixedN{} vary by setup or run.
For example, Opus resolves \Anyo\% at least once and \Everyo\% in all three runs. These are comparisons
of model--harness setups, not isolated models; the primary intervals overlap. Cost means exclude missing
records and interruptions and describe API-price equivalents rather than subscription bills.

\section{Research Opportunities}
\label{sec:rq}

\textbf{Oracle adequacy.} Developer tests can accept incorrect
patches~\cite{yu2025utboost,wang2026solvedissues} and can reject valid ones. The agent patches, per-test
verdicts, and developers' fixes allow both errors to be measured for robot software and compared with
behavioral oracles that drive a simulated robot~\cite{tuzikov2026robobugbench}.

\textbf{Reproducibility of robotics builds.} Recorded build outcomes and environment exclusions
support studies of dependency drift and load-sensitive tests. Postponements must be distinguished from
technical build failures.

\textbf{Issue quality.} Statements are unedited issues, and the audit flag marks tasks whose tests need
identifiers that the issue never names; this supports work on under-specification and issue rewriting.

\textbf{Agent behavior.} The \TrajAvail{} available trajectories from \PrimaryAttempts{} primary attempts support future
analysis of fault localization, build-system use, and domain reasoning; failure causes have not been labeled.

\textbf{Training, test generation, and contamination.} The tasks can serve as training or held-out
data~\cite{pan2025swegym}, their F2P tests as targets for test generation~\cite{mundler2024swtbench}, and
their merge dates for contamination studies.

\textbf{Robotics.} Per-subsystem results motivate testing whether unit-test verdicts predict behavior when repaired
code is replayed in simulation.

\section{Related Work}
\label{sec:related}

\textbf{Issue resolution and reproducible bugs.} SWE-bench~\cite{jimenez2024swebench} defines issue-to-patch
tasks; Verified~\cite{chowdhury2024verified} adds human checks, Multi-SWE-bench~\cite{zan2025multiswe}
extends language coverage, and SWE-bench-Live~\cite{zhang2025swebenchlive} and
SWE-rebench~\cite{badertdinov2025swerebench} automate collection. Defects4J~\cite{just2014defects4j},
BugSwarm~\cite{tomassi2019bugswarm}, and GitBug-Java~\cite{silva2024gitbugjava} package reproducible faults.
ROBUST~\cite{timperley2024robust} provides 221 ROS bugs. \toolname{} combines mined issues, withheld
developer tests, and executable environments across four domains.

\textbf{Robotics coding.} Code as Policies~\cite{liang2023cap}, CaP-X~\cite{fu2026capx},
EmbodiedSWE~\cite{shen2026embodiedswe}, and RLE-Bench~\cite{ma2026rlebench} evaluate newly written control
or learning programs; ROSDevEval~\cite{zhao2027rosdeveval} evaluates function generation.
RoboBugBench~\cite{tuzikov2026robobugbench} evaluates 33 repairs in one ROS~2 navigation stack with
simulation-based behavioral oracles, complementing our inherited tests across \BenchRepos{} repositories.
HWE-Bench~\cite{cui2026hwebench} applies mined fail-to-pass tasks to hardware description code.

\section{Limitations and Future Work}

\subsection{Limitations}
\label{sec:limits}

\textbf{Fairness.} Execution establishes reproducibility, not alignment between an issue and its tests.
An automated audit flags \HiddenN{} tasks whose tests use a name absent from the issue; a patch that
addresses the issue can fail such a test, so the resolve rates are provisional. Human verification, under a two-reviewer protocol modeled on
SWE-bench Verified~\cite{chowdhury2024verified}, has not started; we therefore do not call the dataset
human-verified.

\textbf{Oracle strength.} Containerized tests do not guarantee patch correctness or robot
behavior~\cite{yu2025utboost,wang2026solvedissues}, and resolved patches have not been systematically
audited.

\textbf{Concentration and selection.} Five repositories hold \TopFiveN{} tasks (\TopFivePct\%), mjlab
provides \MjlabN{} of the \BenchDn{} robot-learning tasks, and \SingleTaskRepos{} repositories contribute a
single task, so domain-level results reflect few projects.
The LLM screening has not been audited against human labels.

\subsection{Future Work}

\textbf{Human verification.} Our first priority is to review every task. Reviewers will check that the
developers' fix addresses the issue, that the hidden tests measure the requested behavior, and
that the issue gives an agent enough information without revealing the fix. Tasks that fail will be
removed or excluded from human-verified results.

\textbf{More repositories.} We will relax repository thresholds for activity, merged PRs, and contributors,
and measure the effect on task yield and review outcomes.

\textbf{GPU-dependent projects.} We will support projects such as NVIDIA Isaac Sim with pinned GPU
environments.

\textbf{Simulation-based oracles.} We will add oracles that replay patched code in simulation; they remain
subject to the sim-to-real gap~\cite{josifovski2024cdr,malmir2026diarel}.

\textbf{PRs without issues or tests.} The current dataset requires both a linked issue and developer tests.
For PRs that lack one of them, we will generate the missing part. A generated test must fail before the
developer fix and pass after it~\cite{haratian2026praware}, and human reviewers must confirm that the
issue, implementation, and tests agree and that the issue does not reveal the fix.

\section*{Data Availability}

Dataset: \dataurl. Leaderboard: \toolurl. The archive includes the task directories; the
issue--PR pairs of accepted tasks and rejected candidates, with rejection reasons; agent execution results,
patches, and trajectories; and census data. The Evalium pipeline
source code is proprietary and is not part of the release. Each task keeps its repository's
license: 25 repositories use Apache-2.0, BSD, or MIT, Navigation2 adds LGPL-2.1 packages, and two use
GPL-3.0 (8 tasks), one of which reserves commercial use.

\section*{Acknowledgment}
This work was supported in part by Evalium, which provided computational resources and covered
project-related costs.

\end{document}